\documentclass[letterpaper, 10 pt, conference]{ieeeconf}  

\IEEEoverridecommandlockouts                              
\usepackage[utf8]{inputenc}
\usepackage[T1]{fontenc}

\usepackage{graphicx} 
\usepackage{hyperref}	
\usepackage[table,xcdraw]{xcolor}
\usepackage{placeins}
\usepackage{tabularx}

\title{\LARGE \bf
OSINT Analysis of the TOR Foundation
}

\author{Maxence Delong, Eric Filiol\thanks{Contact author: \url{filiol@esiea.fr}}, Clément Coddet, Olivier Fatou, Clément Suhard\\
        ESIEA Laval, Operational Cryptology and Virology Laboratory $(C + V)^O$ \\ 38 rue des Drs Calmette et Gu\'erin 53000 Laval France}

\begin{document}

\maketitle
\thispagestyle{empty}
\pagestyle{empty}

\begin{abstract}

This research paper is a study on the TOR Foundation and the TOR Network. Due to the blind confidence over this network and this foundation we have collected data and gathered open source information to analyze and to understand how this foundation is organized. We have discovered that the US government is very active through the financial and development aspect. Furthermore, we have discovered that a few critical points about this network, especially that it is managed by unknown people who run and manage special nodes, a fact the foundation never talks about.

\end{abstract}

\section{INTRODUCTION}

The TOR Project (http://www.torproject.org) is the most famous project about privacy and anonymity online. Simply, it is another network where everybody has only the knowledge of his direct neighbor. From this very simple principle, the TOR network routes packets through at least three nodes that ensure that the sender does not know the receiver and conversely. In this network, you can access hidden services which are web service. These services are not indexed by classical search engines. These services are used by several kinds of people and some of them are using the TOR network (and the services provided) because it is their only way to communicate in a safe and anonymous way.

In a country where there is a lot of censorship, journalists can use the TOR network in order to bypass it. In the official website of the TOR project, it is claimed that the TOR network is also used by the military to “protect their communications”. From those perspectives, over the years people have developed a big confidence in the TOR network and in the foundation which controls and manages it. Officially. Allegedly, this foundation claims to have no link with US government (any other one) and is independent (Dingledine, 2017).  There is a growing feeling that this may not be the case. Recurrent questions arise that put this apparent independence into question: what if the US government was behind the TOR network and if somehow controls it? The recent dismantling of Silk road 2.0 by the FBI (Leinwand Leger, 2014) is still not explained from a technical point of view and growing suspicions suggest that a partial control at least by the US government over the foundation cannot be discarded. Hence, if true, the protection of the communication will no longer be true and the confidence in this network should not be warranted. In fact, the TOR project is an implementation of a concept born in the US Naval Research Laboratory (Goldschlag et al., 1996; Syverson et al., 1997). Paul Syverson is the designer of the routing protocol and was part of the original development team of the TOR network. Hence the TOR infancy was clearly linked with the US government and still is.

In this paper, we present the results of a deep OSINT and technical analysis of the TOR network and the Tor foundation which tends to prove that their alleged independence is far from being so clear and acceptable. The paper is organized as follows. We first explain and analyze in the second section what are the TOR directory authorities, what their role is and how they are managed. In the third section,  we explore and expose the internal organization of the TOR foundation and which projects are related to it. In the fourth section we present the results of our deep financial analysis of the TOR foundation funding. It clearly shows that the alleged independence is actually not true. Finally, before concluding we present in the fifth section a few technical aspects that strongly mitigate the claim that the TOR foundation management is suitable for a maximal security and anonymity of the TOR network. All resources presented in the present paper (high definition graphs especially) as well as many others are available on the Blog laboratory (Filiol et al. 2017).

\section{Directory Authority}

\subsection{What is the goal of this node ?}

In this part, we will talk about the directory authorities (see \url{https://svn.torproject.org/svn/tor/tags/tor-0_2_1_4_alpha/doc/spec/dir-spec.txt} for details). The purpose of these specific nodes is to handle the complete network to maintain a list of “currently-running” nodes (this list is refreshed every thirty minutes, so it is not really “currently-running”). All those running nodes are stored in a public file named “consensus”. This file is distributed in order to tell all nodes what is the state of the network, which nodes are running or not.

Every thirty minutes, any node on the TOR network needs to register itself at one or more directory authorities. If a node becomes down or has an issue, it will not be able to register itself on a directory authority. Once each node has been registered, the directory authority will vote to create a new consensus file. The vote is the action of defining the consensus weight of each node in function of the advertised bandwidth, the real bandwidth and the date/time since the node is registered. The consensus weight is one of the most important information in the consensus file because it will define the new bandwidth assigned to each node. As we prove on other research paper (Filiol et al., 2018), the bandwidth is the most important key parameter of the TOR network because we can target specific nodes to take the control of a major part of the traffic. Those nodes (Filiol et al., 2017) are the principal weakness of the TOR network because they define the future state of the network. If we shut down all those nodes, the TOR network will continue to work but will be less and less efficient due to the changes which cannot be specified to the other nodes without the consensus file.  

Currently, there are nine directory authorities, eight for all the nodes and one especially for the bridges (called \textit{Bridge Authority}). Bridges are non-public nodes whose purpose is mostly to enable censorship bypassing. Only the TOR foundation does know the exact list. These directory authorities (bridges authority included) are hard-written in the source code with their IP address, their port and their fingerprint. The concern is that we do not know who is responsible and manages these nodes. No information is available on the website of the TOR Project and the only information available is in the \textit{Changelog} file, but we only have derived information about the node and not the true owner of the node. The purpose of our study is to see the different changes of directory authorities through versions and which is the actual owner of each node. We intended to see the link between owners of directory authority and the TOR Project and how they can justify the possession of specific nodes.

\subsection{Evolution of directory authorities through version}

To do this part of our study, we download all versions of the TOR Project since the beginning. The last version we study is the version \textbf{tor-0.3.0.3-alpha} released on February 3rd, 2017. No changes were made between the last version and the version \textbf{tor-0.3.2.1-alpha}. We have used the Maltego Software to do a graph of directory authorities for each version. The graph is given in Figure 1. The big nodes are the directory authorities which were used for most of the versions. With this classification, we will see which directory authorities are the more relevant to study.

\begin{figure}[h!]
	\includegraphics[width=\columnwidth]{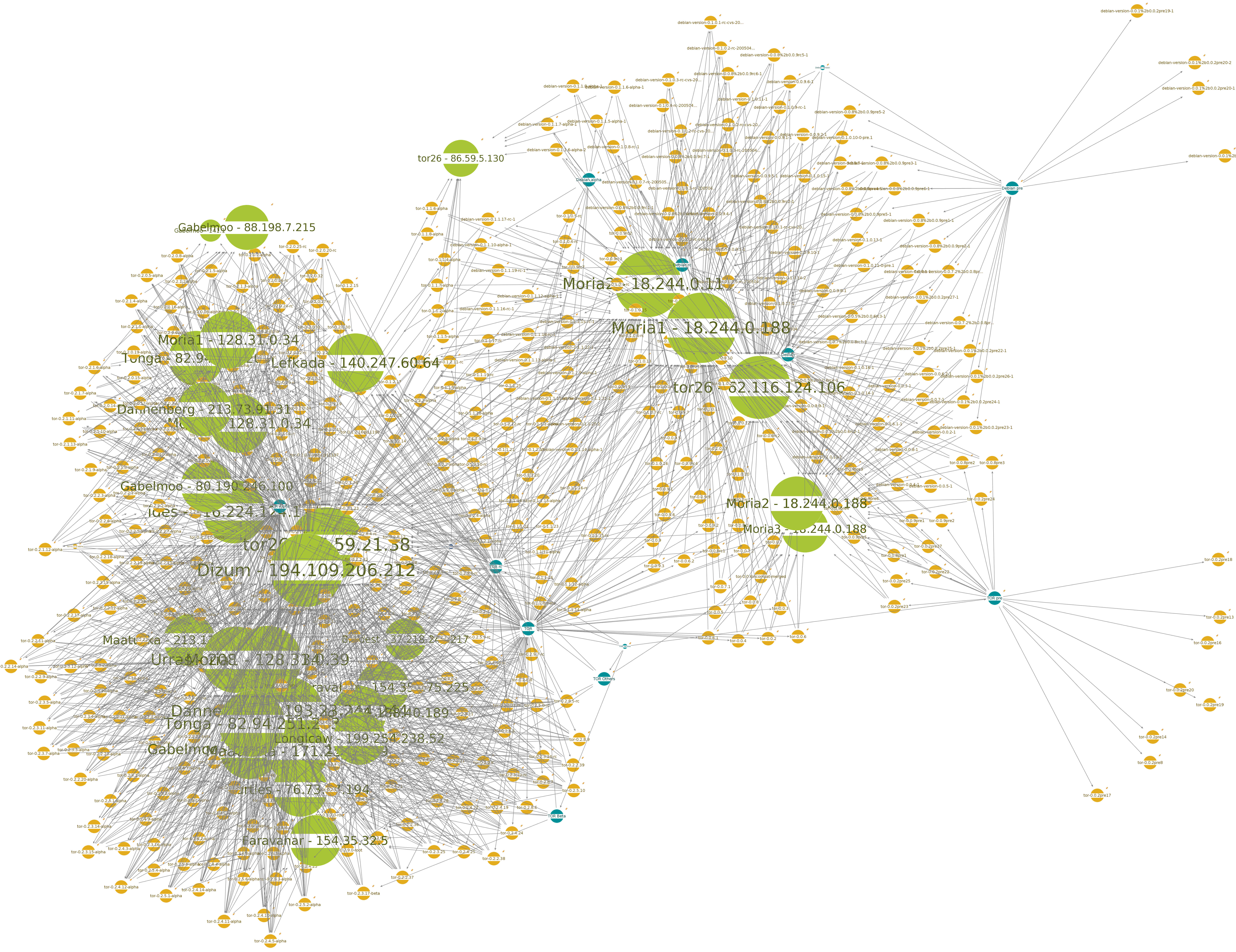}
	\caption{Directory Authorities through TOR version}
\end{figure}

From the graph, we can explain major changes of Directory Authorities. The changes of IP addresses or ports will not be detailed (except if it is the last changes and the most important one). In total, 30 directory authorities were present in different versions (changes of IP addresses included) for 16 different names. In the graph, all biggest points are directory authorities and the smaller ones are each version in the TOR source code.

At the beginning, all directory authorities (three at the time) were owned by the same person on the same IP address but on different ports (Moria1, Moria2 and Moria3). At this time, only Moria1 is still active. Since 2006, one (still active) directory authority was released every two years (except for 2008, we have two directory authorities). In order of appearance, we have Dizum (tor-0.1.1.18, April 2006), Gabelmoo (tor-0.2.0.16, January 2008), Dannenberg (tor-0.2.0.18, January 2008), Maatuska (tor-0.2.2.14, July 2010), Faravahar (tor-0.2.3.23, October 2012) and Longclaw (tor-0.2.6.2, December 2014). In total, there is seven directory authorities and Moria1 still active with the bridge authority. The bridge authority has changed over time. At first, Tonga appears in the version tor-0.2.0.6 in August 2007 to be replaced by Bifroest in the version tor-0.2.9.2 in August 2016. Aside all of these directory authorities, five others directory authorities came and left few versions after. Since the last change (Tonga to Bifroest), no other changes were made in the current version (tor-0.3.2.1, September 2017). Now, let us focus on the nine directory authorities which are present in the last version. Table 1 summarizes how the different versions of the TOR network are using all those directory authorities.

\begin{table}[h!]
	\centering
	\begin{tabular}{|c|c|c|}
		\hline
		& \textbf{Number of versions} & \textbf{\% (456 in total)} \\ \hline
		Moria 1 & 435 & 95\% \\ \hline
		Dizum & 270 & 59\% \\ \hline
		Gabelmoo & 226 & 49\% \\ \hline
		Dannenberg & 223 & 49\% \\ \hline
		Maatuska & 166 & 36\% \\ \hline
		Faravahar & 110 & 24\% \\ \hline
		Longclaw & 74 & 16\% \\ \hline
		Bifroest & 47 & 10\% \\ \hline
	\end{tabular}
	\caption{Number of version using those directory authorities}
\end{table}

As we can see, a few of the directory authorities are very recent. Taling into account the fact that the versions of TOR are released at least three per three (multiple versions are changed in the same time), the array proves that the TOR Project puts a high confidence in their owner. However, up to us, the user does not have any information about the owner of those very important nodes. A study of each user, their connection with the TOR Project is necessary to evaluate the actual confidence users can give.
\subsection{Owners of directory authorities}
For every owner of directory authorities, we have made an OSINT study on this people to see what are their connections with the TOR Project and if one can actually trust these people. In addition, we have conducted a study on the nodes themselves including geolocation and different information on those nodes. We consider as a “confident people” the person that has a visible connection to the TOR Project (through their “Core People” pages, through projects around TOR etc\ldots). We will also study the directory authority one by one, starting with the most used.

For \textit{Moria1}, the current IP address is 128.31.0.39. Since the first version, this node is on one of the two MIT networks (128.31.0.0/16 or 18.238.0.0/17). This node is geolocated in Cambridge (Massachusetts – United States). The owner is Roger Dingledine, one of the three creators of the TOR Project (and a former NSA employee). At the time of the study, ten “classic” nodes were registered in the MIT network. We were able to gather those nodes either with their IP address or with their hostname (ending by mit.edu).

The second most important node is \textit{Dizum}. This directory authority has never changed (IP address 194.109.206.212). The geolocation points towards the Netherlands, near Dordrecht. His owner is Alex De Joode and we were not able to find any link between Alex De Joode and the TOR Project. Is it annoying when we see that his directory authority is the most important one. In fact, a full blind confidence to this person is given while it is not possible to link him to the TOR Project in a way or another.

The third is \textit{Gabelmoo} on the IP address 131.188.40.189 (on the network 131.188.0.0/16). This directory authority is the most mobile of all, with five changes during almost ten years. For the last IP address, the geolocation indicates Erlangen in Germany. This directory authority is set up on a university network. In this network, three “classic” nodes are registered. The owner is Sebastian Hanh who is a developer on the TOR Network. Another node belongs to Sebastian Hanh but on a different network (IP address 78.47.18.110).

The fourth is \textit{Dannenberg} (193.23.244.244) and has changed once. This modification does not change the geolocation of this node, which is still in Berlin. This node is hosted in the Chaos Computer Club (CCC) and handled by Andreas Lehner. He is registered on the Core People page.

For \textit{Maatuska} (171.25.193.9), only one change of IP address was made (old IP address: 213.115.239.118). This directory authority is apparently located in Stockholm in Sweden (RIPE Database). In the range 171.25.193.0/24, eight “classic” nodes are set up. Those nodes are handled by Linus Goldberg who is a confirmed member of the TOR Project.

\textit{Faravahar} (154.35.175.225) has changed only once (previous IP address 154.35.32.5). Sina Rabbani is the handler of this directory authority. He works with the TOR Project and is present on the Core People page. The node is located in the USA (Washigton DC according to the ARIN Database). No other “classic” nodes are on the same network. 

The last directory authority (excluding the bridge authority) is \textit{Longclaw} (199.254.238.52). It has been set up in 2014 and this node is too recent to have previous changes. The owner of this directory authority and three “classic” nodes (on the network 199.254.238.0/24) is the foundation \textit{Riseup Network}. The geolocation is not very precise: this node is indicated as located in Missouri, USA. With our study, we can affirm that Micah Anderson is behind the RiseUp Network and is the real owner of this directory authority. He has recently disappeared from the Core People page.

The last one and the most important node authority is the bridge authority \textit{Bifroest} (37.218.247.217). \textit{Bifroest} is not the first bridge authority. \textit{Tonga} was in place before leaving the place to \textit{Bifroest}. This bridge authority is in place in Amsterdam (Netherlands). The person known as Isis Agora Lovecruft is responsible of the bridge authority. However, this is not the person who is registered in the RIPE Database but two members of the GreenHost firm instead.

To conclude, we see that seven people over the nine have a visible relation to the TOR Project. This means, we have let the control of the TOR network to two unknown people (Micah Anderson and Alex De Joode). Alex De Joode is the oldest directory authority which different from those belonging to Roger Dingledine. From the beginning, he has helped running the TOR network but we do not have any information about this person. It is the same situation for Micah Anderson who is hidden behind the RiseUp Network. It is a real problem for the network: do users can trust people they do not know? Where do these people come from? What is their background?

\section{Project Participants and internal organization}

During our study, we have gathered different files containing a list of people who have participated in the TOR Project. Our purpose is to explain, in a simple way, the organization inside the TOR Project Inc. (Name of the company). If the TOR Project is really transparent in this internal organization, we should be able to have a list of all participants. This list is available in the Core People page (\url{https://www.torproject.org/about/corepeople.html.en}). Furthermore, a few personal repositories were opened and we have downloaded the data in order to learn more about this firm. We will present our study results about those people, their role and expose a few hidden people.

\subsection{Core People} 
We have collected information about every person who is present in the official website of TOR from the beginning of 2017. The website has changed since and we have seen that a few people have disappeared and other have appeared. We have divided those people in seven groups: Founders, Researcher, Developer, Advocate, Administrative ant Tech leader, Technical Support and Others (see Figure 2). In this web page, we have information regarding the role of each people in the TOR Project. 

\begin{figure}[h!]
	\includegraphics[width=\columnwidth]{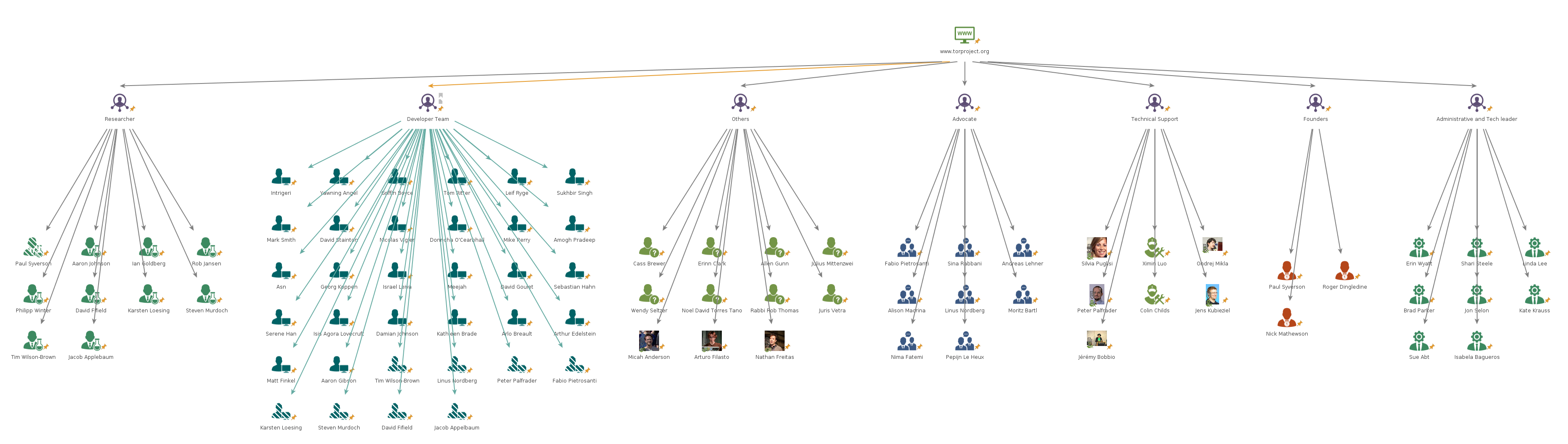}
	\caption{Core People}
\end{figure}

First, we focus on the career path of the three founders. The founders are the developers of the first version of the TOR protocol. Nick Mathewson is a former MIT student as Roger Dingledine was. With the first directory authorities \textit{Moria1, Moria2} and \textit{Moria3} hosted in the MIT, we can say that the MIT is the location where the first implementation of the onion routing was set up (the very first version of the TOR network). The third person is Paul Syverson which is a US military from the US Naval Researcher Laboratory. He and his team (in the US NRL) have designed the onion routing protocol (Goldschlag et al., 1996; Syverson et al., 1997). The TOR project is only the visible part of a larger US project.

After having created the TOR network, Paul Syverson has continued to work with the TOR foundation. There are at least 25 research papers coming from Paul Syverson for the TOR network. The last example in date was the 18th of September 2017 for the version tor-0.3.2.1 which was implemented by following a paper wrote by Paul Syverson and his team from the US NRL only. Furthermore, Roger Dingledine spent a summer in internship in the NSA, so we can suppose that he has kept a few contacts in there. Officially, TOR is not developed anymore by the US government but a major part of changes was designed and developed by Paul Syverson through the US NRL and some people have work closely for the US government (not only among founders). 

With the change of the Core People page, a few people have disappeared. A lot of them were simply developers (6 in total), two were from the technical support and one was a researcher. The interesting person who has disappeared is Jérémy Bobbio, better known as \textit{Lunar}. His repository was huge and thanks to him, we have discovered all the presentations done about the TOR project. In this conference, we found a number of presentations to the highest authorities of the United States. Two presentations of the project were made in the FBI on 2009 and 2012, one in the Naval Criminal Investigative Service (NCIS) which is the investigation unit affiliated to the US Navy (as the US NRL) in 2012 and one in the White House in 2013. 

Two other people have disappeared: Micah Anderson and Peter Palfrader. Those two people were (or are) owner of a directory authority. Peter Palfrader was the owner of \textit{tor26} (the first directory authority which does not belong to Roger Dingledine). Released in the version tor-0.0.8.1 in October 2004, the directory authority is not working anymore. It is more important for Micah Anderson because we do not longer have information on the website. Previously, on the old version of the Core People page, we have his nickname and a little description explaining that he ran one of the directory authorities. If a few people need to be on the Core People page, it will be the founder of the TOR Foundation and the people running a directory authority. With this disappearance, the customers have less information about the people who actually handle the network.

We note that the Core People page is not containing information about a few important people in the TOR Foundation. This page is not sufficient to have an idea of who are the true leaders of the foundation. We have explained who are the leaders of the network (directory authorities) but not those of the foundation.
\subsection{Employees}
To get a list of all employees and leaders (administratively speaking), we have used the form 990. This form is originally for the accounting section (cf 4.1 Form 990 – Global Revenue 2015) but there is a list of all paid employees. This form is not released for the year 2016 or 2017 so we have study the document from 2015.

In the part VII, we have the list of directors, key employees\ldots We discover a few names that are not on the Core People page. Rob Thomas, Meredith Dunn, Andrew Lewman, Mike Perry and Andrea Shepard are still unknown. They represent five over eleven people (~45\%) that are key people for the foundation. Some contractors were hired, Pearl Crescent for example (a developer), and were “hidden” by the foundation. The TOR foundation asks indirectly a blind trust on the source code (due to the huge amount of line) and they give the development to people we do not even know. 

Due to the scandal with Jacob Appelbaum (Greenberg, 2016), the board of directors has completely changed. Every director, including Roger Dingledine and Nick Mathweson has left to let the place to six new board members. Among them we find Matt Blaze, Cindy Cohn, Gabriella Coleman, Linus Nordberg, Megan Price and Bruce Schneier. In this list, Megan Price and Bruce Schneier are not present on the Core People page. To draw a hierarchical graph of the foundation, we need to wait the form 990 for the year 2016 (which is still not available). It is a big change for the foundation. As we said, we have worked on the previous Core People page and the form 990 from 2015 and no one (except Linus Nordberg) was known. 

No relevant information is available on the webpage, so we have looked for others information in different other places and with different other sources. Matt Blaze is an Associate Professor of Computer and Information Science in the University of Pennsylvania. He has no previous link with the TOR Foundation. Cindy Cohn is a Lawyer specialized in Internet laws. She is the current Executive Director of the Electronic Frontier Foundation (EFF). She worked on the case Hepting v. At\&T which is a case where AT\&T was accused to help the NSA to collect data. The case was dismissed on June 2009. She has no previous link with the TOR foundation. Gabriella Coleman is an anthropologist who studies hacking and online activism. She is in the board of several organizations to defend the women rights and private life. As the two-previous people, she does not seem to have previous history with the TOR foundation. Megan Price is the Executive Director of the Human Rights Data Analysis Group (HRDAG). She has no link with the TOR foundation before joining the board. Bruce Schneier is an American cryptologist. Before creating his company (Counterpane), he has worked with the Department of Defense (DoJ). He has no previous link to the TOR foundation. Linus Goldberg is the only board member who has worked on the TOR network before.
\section{Financial Analysis}
The TOR Foundation is regularly claiming that the US government is not funding anymore the TOR Project (Dingledine, 2017). To confirm or not this affirmation, we have analyzed two different documents which are present on the TOR Project website. The first is the form 990 that we used for gathering the list of employees. The second is the “\textit{Consolidated Financial Statements and Reports required for Audits in Accordance with Government Auditing Standards and Uniform Guidance}”.

The purpose of this study is to understand the position of the US government regarding the TOR network and if whether it is involved in the different funds or not. The position of the US government for the TOR network is odd, strange and weird. In one hand, the US government has authorized the US NRL to publish papers for the onion routing protocol. In the other hand, different organizations like the \textit{Federal Bureau of Investigation} (FBI) or the \textit{Central Intelligence Agency} (CIA) are trying to crack the TOR network. The \textit{National Security Agency} (NSA) is often involved in rumors which denounced spying or cracking the TOR network.

We have proceeded by studying the last financial report at our disposal. The financial statements for 2016 or 2017 are still not released to the date of the present paper so the financial report of 2015 was our only option. At the end of this study, an evolution of the funding of the TOR project will be made.
\subsection{Form 990 - Global revenue - 2015}
First, we study the global revenue of the TOR foundation for the year 2015. For this, there are four categories: Contributions and grants, Program service revenue, Investment income, other revenue. The following graph is the repartition of all incomes.

As we can see in Figure 3, most of the funds are coming from the category “\textit{Program Service}”. The other major part is from the “\textit{Contribution and grants category}”. In this category, there is no obligation to publish the donators and the amount of donation for each. To be more precise on our study, we tried to get more information about the Program service revenue.

\begin{figure}[h!]
	\includegraphics[width=\columnwidth]{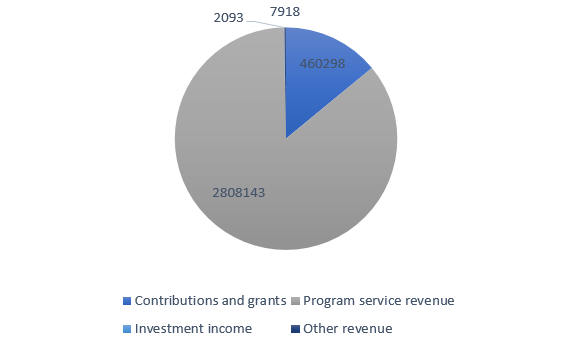}
	\caption{Repartition of Funds}
\end{figure}
\FloatBarrier
\subsection{Program revenue - 2015}
The aim of this study is to determine whether the rumors according to which the US government would help a lot the TOR foundation from a financial point of view, is true or not. For that purpose, we have gathered information from the two above-mentioned sources. In Table 2, we have detailed the entire program and computed the funding proportion of each program.

\begin{table}[h!]
	\centering
	\begin{tabularx}{\columnwidth}{X|X|X|X|X}
		\hline
		\multicolumn{2}{|l|}{} & 2015 - \$ & Details & 2015 - \% \\ \hline
		\multicolumn{2}{|l|}{Program service} & 2,808,143. & 2,808,143. – 100\% & 85,65\% \\ \hline
		\rowcolor[HTML]{9B9B9B} 
		& RFA Contract &  & 886,724. – 31,58\% &  \\ \hline
		\rowcolor[HTML]{9B9B9B} 
		DoS & Direct (DRL grant) &  & 857,515. – 30,54\% &  \\ \hline
		\rowcolor[HTML]{9B9B9B} 
		DoD & SRI Lights Contract &  & 719,500. – 25,62\% &  \\ \hline
		\rowcolor[HTML]{9B9B9B} 
		US Independent Agency & Direct (NSF) &  & 225,184. – 8,02\% &  \\ \hline
		\rowcolor[HTML]{9B9B9B} 
		DoS & Internet Networks &  & 104,540. – 3,72\% &  \\ \hline
		& Unknown &  & 13,500. – 0,48\% &  \\ \hline
		\rowcolor[HTML]{9B9B9B} 
		US Independent Agency & NSF (through RUM) &  & 1,180. – 0,04\% &  \\ \hline
		\multicolumn{2}{|l|}{Contrib \& Grants} & 460,298. &  & 14,04\% \\ \hline
		\multicolumn{2}{|l|}{Other revenue} & \multicolumn{1}{c|}{7,918} & \multicolumn{1}{c|}{} & \multicolumn{1}{c|}{0,24\%} \\ \hline
		\multicolumn{2}{|l|}{Investment income} & \multicolumn{1}{c|}{2,093} & \multicolumn{1}{c|}{} & \multicolumn{1}{c|}{0,06\%} \\ \hline
		\multicolumn{2}{|l|}{Total} & \multicolumn{1}{c|}{3,278,452} & \multicolumn{1}{c|}{} & \multicolumn{1}{c|}{100\%} \\ \hline
	\end{tabularx}
	\caption{Funding Details for the year 2015}
\end{table}

In Table 2, incomes from (directly or not) the US government are underlined in grey. As we can see, at least 58.20\% of the total funds are coming from different departments of the US government. The status of RFA (Radio Free Asia) Contract is unclear and there are persistent allegations and testimonies (Prados, 2017; Levine, 2015) or even suggestions that it could be strongly connected to the CIA more than expected (Levine, 2015). Would this suspicion be true, the rate of funds from US government-related entities would grow up to 85.24\%. With these two-huge ratios, we can say that the US government is very active from the financial point of view (let us note that a few incomes are still unknown because the TOR project does not need to release the list of ALL donators).

\subsection{Funds from the US government since 2007}
The accountings are available since the TOR foundation gets the status of non-profit organization. With all data gathered during our study, we were able to study the financial report of each year and to determine which funds are from the US government directly or indirectly (Figure 4).

\begin{figure}[h!]
	\includegraphics[width=\columnwidth]{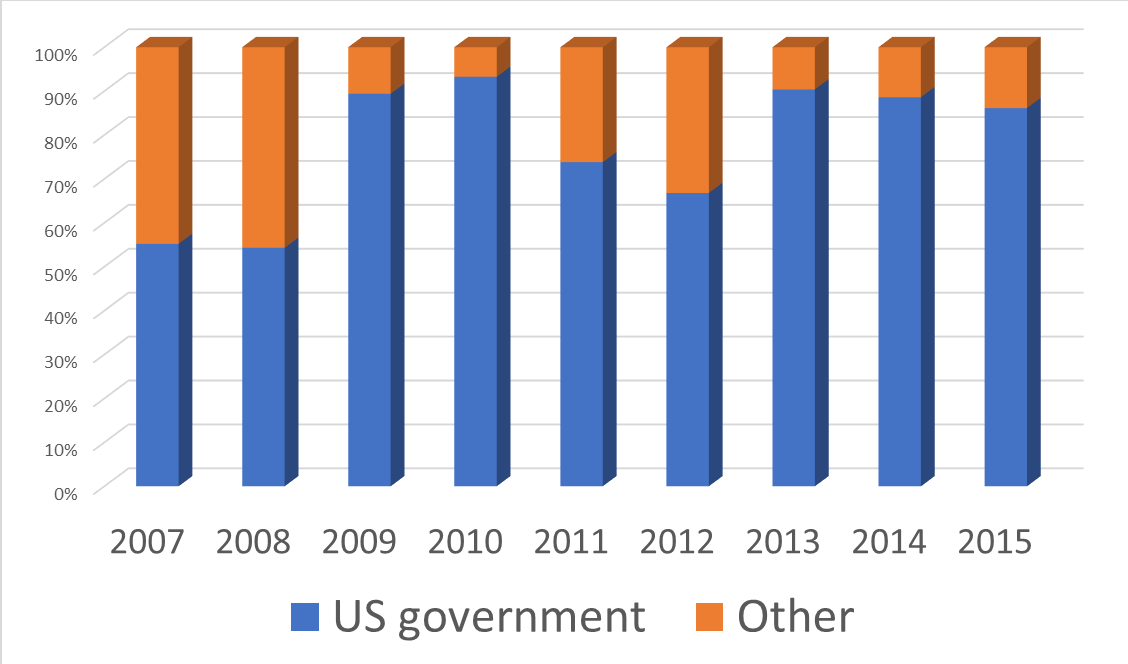}
	\caption{US funding since 2007}
\end{figure}
\FloatBarrier

As explained previously, we are not sure that some incomes are from the US government. To avoid uncertainties, in this graph, we give the minimal value of the US government implication regarding the revenue. By studying all the annual financial reports, we see that since 2011, the list of programs is almost the same (NSF, SRI, Internews and DRL). The other interesting thing is the progression of the total revenue per year. For example, between 2007 and 2008, only 80,000\$ were added to the annual incomes however, between 2011 and 2012, the revenue just doubled. To illustrate the progression, Table 2 summarizes the changes year per year. Years 2009 and 2012 are interesting: the global revenue almost doubled in one year. Each time, it is the “\textit{Program revenue}” category which was multiplied by 2. No event is simply identifiable which would justify such an increase of the revenue.

\begin{table}[h!]
	\centering
	\begin{tabular}{|c|c|}
		\hline
		& \textbf{Percentage increase} \\ \hline
		2007 $\Rightarrow$ 2008 & 17.313 \% \\ \hline
		2008 $\Rightarrow$ 2009 & 96.126 \% \\ \hline
		2009 $\Rightarrow$ 2010 & 28.290 \% \\ \hline
		2010 $\Rightarrow$ 2011 & 3.797 \% \\ \hline
		2011 $\Rightarrow$ 2012 & 88.088 \% \\ \hline
		2012 $\Rightarrow$ 2013 & 10.121 \% \\ \hline
		2013 $\Rightarrow$ 2014 & -11.018 \% \\ \hline
		2014 $\Rightarrow$ 2015 & 28.246 \% \\ \hline
	\end{tabular}
	\caption{Augmentation of revenue since 2007}
\end{table}

\section{Technical Aspects}
We will not develop most of the technical aspects that could suggest or confirm that somehow the TOR network has been designed or is managed in such a way that a few “facilities” are possible and would enable to take control over it. In (Filiol and al., 2017 \& 2018) it has been proved that the TOR routing protocol suffers from rather severe flaws. As a consequence, taking the control of a reduced number of TOR relays (from 450 to 1400 only) would enable to reduce the TOR traffic of at least 50 \% and would greatly ease correlation attacks (about 35 \% of the traffic) or eavesdropping (about 10 \% of the traffic). 

As far as the relay bridges management is concerned, it has been possible to extract slightly more than 2,500 such bridges thus compromising the alleged ability to bypass censorship. Following the disclosure of our results (Filiol and al., 2017) we got a very interesting feedback and case that strongly confirms that security management in the Tor network is not sufficient and thus intelligence entities could easily do really undercover things.

During our study, in September 2017, we were contacted by a user of a custom TOR library. This library is the “node-Tor” written in JavaScript and allows the user to create and run a node or connect to the TOR network. Further exchanges with this person have shown a lot of inconsistency and irregularities.

At first, we talk about the way that his node was added to the network. For this custom library, the user asked the TOR foundation to add a node with this library and after an exchange of a few mails the node was accepted and run. The library is very different from the original source code. To compare very simply those two codes, we just compared the number of code lines. We know that the number of code lines does not really reflect the effect of the code but between the original source code (several hundreds of thousands code lines) and the library (only fifteen hundred code lines), we can assure that it is very likely that a number of options or securities are missing.

It is not the designer of this code who is responsible but rather the TOR foundation for accepting a node on the network with this kind of library. The first problem is that no one is warned that this node is special and is not running the official source code. This node owned by a user is not controlled by the TOR foundation. So if the user is malicious, he could modify his node and make every change he wants. If a government wants to include this kind of node to log the traffic and gather it, he can do it very simply and without triggering any alert.

For now, his node is running in two ports: one on the port 8001 for the classical connection and on the port 8002 for a websocket port. To use the TOR network, every client needs to create a circuit of three nodes. The circuit is created according to the following process:
\begin{itemize}
	\item Pick randomly three nodes (in the case of external circuits)
	\item Send “Create Path” to the guard node
	\item Send “Extend Path” to the guard
	\item Send “Extend Path” to the middle
\end{itemize}
In the case where a custom node-TOR is used as guard or middle node, it will not be able to extend the circuit. But, if the node is used as an exit node, the node can be used because it will not receive an “Extend Path” frame. When we talked to the owner of this node, he ensured us that the node is not able to extend and that node was here only for research purposes but in reality, this node can be used as an exit node.

By checking his log, he noticed that circuits were created whenever using the TOR Browser. One of the circuits was his own node-TOR on port 8002 for the entry node, a random TOR node for the middle node and his node-TOR on port 8001. Here, we have a serious problem regarding the family policy established in the TOR protocol (in the \textit{torrc} file precisely). After checking the node-TOR library, we were not able to find the management of the family. This proves that the TOR foundation is accepting every library which looks like the original source code. Among the users of the TOR network, a few of them only have really read the source. As they do not read the source code, they have a blind trust of this network and the TOR foundation accepts any custom library. If the security of the network is ensured by the fact that all the nodes run the same source code, with the same security level, the same options and so on\ldots this fact proves that the TOR network is not so secure.  

Then, we discussed about how the node was accepted on the network. For running his node with the library, he was asked to register his node on several directory authorities (\textit{Gabelmoo, Dizum, Tonga, Tor26, Dannenberg, Maatuska, Urras, Turlte} and \textit{Moria1}). \textit{Tonga} is the ancestor of \textit{Bifroest}, the bridge authority. Without knowing it, his node was not registered on the “classical” consensus file but in the consensus for the bridge. Against his will, his node was considered as a bridge during all the time and he was not even aware of this. He became aware just when we released the list of 2,500 bridges.

Today, the logs are different: almost all directory authorities do not register this custom node anymore, except one: \textit{Urras}. \textit{Urras} is no longer a directory authority since 2015. This node belongs to Jacob Applebaum. It means that apparently, he continues his role at least with this custom node. However Jacob Applebaum has been fired from the TOR foundation and is no longer supposed to keep this particular role.

This feedback from this user was really interesting. We have discovered that with only few exchanges with the TOR foundation, we can add a custom node (possibly malicious). As for every node, no systematic control is possible by the TOR foundation, once accepted within in the network, we can do what we want with this node, log the traffic, insert biases in the creation of circuits etc\ldots In summary, we think that the TOR project should not accept custom codes in order to respect the uniformity of the network that ensures “security”.  

\section{Conclusion}
To conclude this presentation of our study, let us make a summary of all information that we have gathered and conclusions we have drawn. First, the critical point of the TOR network: the directory authorities. As we saw, some of them are handled by “unknown” people. In the real world, we teach children not to speak with unknown people but here, it seems not disturbing for most of the TOR users to trust perfect strangers. As far as confidence is concerned, nobody (except state organization) has the courage/time to read the source code and no one is paying attention to the designer of the changes on the TOR source code. Paul Syverson (from the US NRL) is the original designer (not developer) of most of implementation. The last version of TOR is the perfect example: all major changes are coming from the US NRL. The US government is not just present on the source code: he is also present on the funding of the TOR project. Some years, more than 80\% was given by the US government through various Departments (State and Defense mostly).
 
No official statement revels that the US government is helping the TOR network but all the information gathered during our study seems to confirm that the US government is still deeply involved in the TOR project. The privacy online and the anonymity are more and more present in every debate about Internet. The TOR project has recently announced that they provide those characteristics (Dingledine, 2017) and many peoples rely on this and trust them. For journalists in a country where the censorship is very present, for political activist the TOR network is the only solution to hide themselves and protect their real identity. For the militaries that use TOR to communicate in a secure way, if a country, an organization has the control of the channel or a way to spy the communication, it would be a disaster.

This study is not claiming breaking the TOR network or affirms that the US government is the real organization behind the TOR project. However favoring such a network would be a clear violation of the Wassenaar Agreement (\url{www.wassenaar.org}) unless some sort of control is in place in a way or another (Filiol, 2013). This study aims at informing TOR users and to make them aware of network like the TOR network and the possible reality behind. Customers need to be informed before using any network who claims to protect your privacy and anonymity.

\addtolength{\textheight}{-12cm}   




\end{document}